\documentclass{article}
\usepackage{smc2026}
\usepackage[caption=false, font=footnotesize]{subfig}
\usepackage{paralist}
\usepackage[figure,table]{hypcap}
\usepackage[english]{babel}
\usepackage{amsmath}
\usepackage{booktabs}
\usepackage{multirow}

\def\papertitle{Acoustic Overspecification \\\ in Electronic Dance Music Taxonomy}

\author[1]{\mbox{\firstname{Weilun}\lastname{Xu}$^\dagger$}}
\author[1]
{\mbox{\firstname{Tianhao}\lastname{Dai}}}
\author[1]{\mbox{\firstname{Oscar}\lastname{Goudet}}}
\author[2]{\mbox{\firstname{Xiaoxuan}\lastname{Wang}}}

\affil[1]{\department{School of Computer and Communication Sciences}\institution{École Polytechnique Fédérale de Lausanne}\country{Switzerland}\affiliationtype{University}}
\affil[2]{\department{Digital and Cognitive Musicology Lab}\institution{École Polytechnique Fédérale de Lausanne}\country{Switzerland}\affiliationtype{University}}

\completesetup

\title{\papertitle}

\begin{document}
	\capstartfalse
	\maketitle
	\capstarttrue
	\renewcommand{\thefootnote}{\fnsymbol{footnote}}
	\footnotetext[2]{Corresponding author. The code and supplementary material are available at \url{https://github.com/alunxu/edm-overspec}.}
	\renewcommand{\thefootnote}{\arabic{footnote}}

	\begin{abstract}
		Electronic Dance Music (EDM) classification typically relies on industry-defined taxonomies, with current supervised approaches naturally assuming the validity of prescribed subgenre labels. However, whether these commercial distinctions reflect genuine acoustic differences remains largely unexplored. In this paper, we propose an unsupervised approach to discover the natural acoustic structure of EDM independent of commercial labels. To address the historical lack of EDM-specific feature design in MIR, we systematically construct a tailored, interpretable acoustic feature space capturing the genre's defining production techniques, spectral textures, and layered rhythmic patterns. To ensure our findings reflect inherent acoustic structure rather than feature engineering artifacts, we validate our clustering against state-of-the-art pre-trained audio embeddings (MERT and CLAP). Across both our bespoke feature space and the pre-trained embeddings, clustering consistently identifies 20 or fewer natural acoustic families---suggesting current commercial EDM taxonomy is acoustically overspecified by nearly one-half.
	\end{abstract}

	\section{Introduction}\label{sec:intro}
	Electronic Dance Music (EDM) is commercially organized into unusually fine-grained subgenre taxonomies whose acoustic basis remains untested. Platforms like Beatport\footnote{\url{https://www.beatport.com} -- leading digital music platform specializing in electronic dance music.} categorize tracks into 35 or more distinct subgenres, yet whether these distinctions reflect genuine acoustic differences or primarily serve commercial and cultural functions remains an open question~\cite{brackett2016}. To understand this tension, we distinguish between two fundamentally different conceptions of genre. \emph{Acoustic genres} are computationally discoverable clusters based on measurable sonic features---tempo, timbre, harmonic content, and rhythmic patterns---reflecting genuine perceptual differences. \emph{Cultural genres}, by contrast, are socially constructed categories that serve marketing differentiation, regional scene identity, and what Thornton terms ``subcultural capital''---knowledge of micro-genres functioning as cultural currency within dance music communities~\cite{thornton2013club}.

	Previous EDM classification research has predominantly employed supervised learning with industry-provided labels~\cite{caparrini2020,shu2024,knees2015,faraldo2016,popli2022,hsu2021}. However, these approaches assume label validity without examining whether the prescribed taxonomy aligns with acoustic reality. This assumption is precisely what we test: genre distinctions in EDM are frequently tied to specific production tools and techniques---from the Roland TB-303 that defined Acid House to the sidechain-compression aesthetics of modern Progressive House~\cite{butler2014playing}---and these production-level cues may not be fully captured by standard audio features, further complicating the relationship between acoustic measurement and genre identity.
        
	\begin{figure}[t]
		\centering
		\includegraphics[width=\columnwidth]{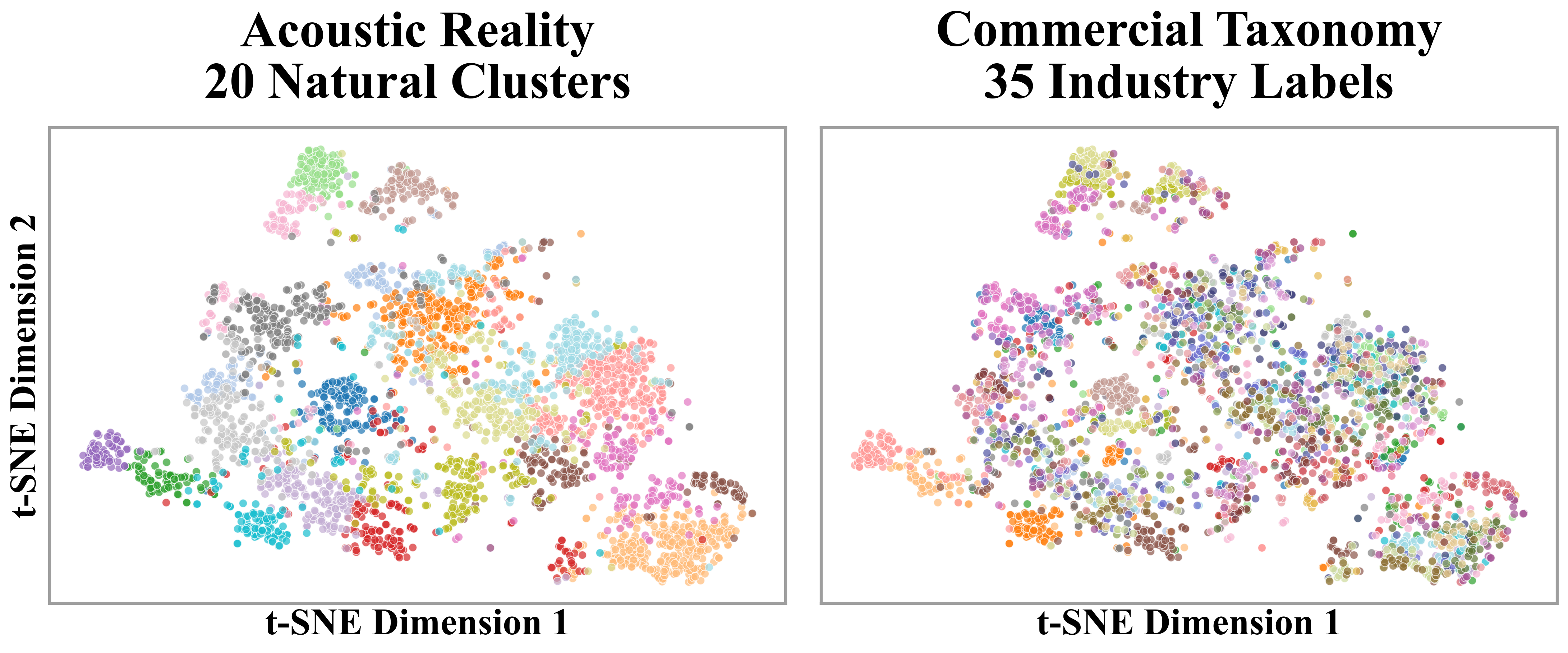}
		\caption{t-SNE projection of selected acoustic features (Section~\ref{ssec:features}). Left: K-means clusters at the optimal $k$ identified in Section~\ref{ssec:natural}. Right: identical projection colored by the 35 commercial Beatport genre labels. The right panel's color fragmentation within spatially coherent regions illustrates the overspecification of industry taxonomy relative to acoustic structure.}
		\label{fig:tsne}
	\end{figure}

	We therefore ask: Does EDM's commercial genre taxonomy reflect its underlying acoustic structure? Figure~\ref{fig:tsne} visualizes this tension---when the same tracks are organized by acoustic similarity versus market labels, different patterns emerge. We investigate this disconnect through clustering, allowing natural groupings to emerge without presupposing the validity of industry categories.

	Our contributions include: (1) An interpretable acoustic feature space tailored to EDM---spanning spectral-harmonic, production, texture, and rhythmic dimensions---which may inform further downstream tasks such as genre discovery, recommendation, and classification; (2) Evidence that EDM's commercial taxonomy is \emph{acoustically} overspecified: unsupervised clustering consistently identifies 17--20 natural acoustic families versus 35 prescribed genres, suggesting that many commercial subdivisions lack acoustic differentiation. This finding concerns acoustic structure only; genre labels serve cultural, historical, and commercial functions that extend well beyond acoustic separability~\cite{brackett2016}.

	\section{Dataset and Features}\label{sec:data}
	
	\subsection{Dataset}\label{ssec:dataset}
	We construct our dataset from Beatport, collecting the top 100 hit tracks from each of 35 genres (3,500 total) as of March 2025. On Beatport, genre labels are assigned by artists and labels who select from a curated taxonomy maintained by the platform. However, this process is not purely bottom-up: Beatport retains editorial oversight and periodically revises these categories. Genres thus originate from community and artist practice but are centrally mediated through Beatport's explicit curation. Each track is listed under a single primary genre, so our dataset contains no duplicate tracks across categories. While broader terms like ``Techno'' qualify multiple Beatport categories (e.g., Hard Techno, Melodic House \& Techno), we treat each strictly as a distinct genre. One category, DJ Tools, is not a traditional genre but a utility class containing loops, samples, and mixing elements; we retain it because it is part of Beatport's official taxonomy, and a sensitivity analysis\footnote{See Appendix~C in the supplementary material, available at \url{https://github.com/alunxu/edm-overspec}.} confirms its inclusion does not materially affect our results. We extract 2-minute previews at 320\,kbps with metadata including BPM, key, length, and genre labels. These chart-topping tracks serve as established genre exemplars.

	\subsection{Feature Extraction}\label{ssec:features}
	We extract 194 acoustic dimensions per track, processing audio identically at 44.1\,kHz. To capture EDM's production techniques, texture, and layered rhythmic patterns, we extract three complementary feature groups (see the supplementary material\footnote{See Appendix~A and Table~1 in the supplementary material.} for a complete taxonomy).

	\textbf{Conventional MIR Features} (92 dimensions): Following established baselines~\cite{caparrini2020}, we extract (1) spectral shape features including centroid, spread, entropy, flux, and rolloff, (2) timbral descriptors comprising 13-band MFCCs, (3) harmonic features via 12-dimensional chroma vectors, and (4) rhythmic measures including autocorrelation-derived danceability, beat histogram periodicity, and frequency-band beat emphasis (energy restricted to beat-frames across 6 octave-spaced bands). We compute the mean and standard deviation of these frame-level metrics.
	
	\textbf{EDM Production Features} (38 dimensions): We adapt standard audio analysis tools alongside novel proxies to capture club-oriented mixing aesthetics. For dynamics, we compute the Harmonic-to-Percussive (H/P) energy ratio, crest factor, and RMS kurtosis using median-filtering separated components~\cite{fitzgerald2010} to quantify drum prominence and dynamic range compression (brickwalling). For spectral texture, we extract 7-band spectral contrast~\cite{jiang2002} to track synthesizer sharpness versus noise. For bass design, we introduce a novel sub-bass ratio (power beneath 60\,Hz) and a sidechain pumping proxy that measures the negative correlation between the kick-band envelope and high-frequency synthesizer envelope.

    \noindent \textbf{Tempogram-based Features} (64 dimensions): 
    Standard tempo estimation often fails to capture the layered rhythmic structures inherent in EDM. Following \cite{grosche2012}, we address this by computing three complementary tempogram representations. 
    
    First, the \textbf{Fourier tempogram} applies a short-time Fourier transform to a novelty curve (the onset-strength envelope). To map standard frequency $\omega$ (in Hz) to musical tempo $\tau$ (in BPM), we use the relation $\tau = 60 \cdot \omega$. The tempogram is defined as:
    \begin{equation}
    	T^F(t,\tau) = |\mathcal{F}(t,\tau/60)|
    \end{equation}
    where $t$ is the time frame and $\mathcal{F}(t,\omega)$ denotes the windowed Fourier transform of the novelty curve. 
    
    To capture subharmonic periodicities, we compute an \textbf{autocorrelation tempogram}. This compares the novelty curve against time-shifted copies of itself using a time lag $\ell$. The tempo $\tau$ relates to this lag via $\tau = 60 / (\ell \cdot r)$, where $r$ is the frame step size in seconds:
    \begin{equation}
    	T^A(t,\tau) = \mathcal{A}\left(t, \frac{60}{\tau \cdot r}\right)
    \end{equation}
    where $\mathcal{A}(t, \ell)$ represents the local autocorrelation of the novelty curve at time $t$ and lag $\ell$.
    
    To remove octave ambiguities within both representations, we derive \textbf{cyclic tempograms} by pooling all tempi related by powers of two into a single octave:
    \begin{equation}
    	C_\rho(t,s) = \sum_{\lambda \in [s\cdot\rho]} T(t,\lambda), \quad s \in [1,2)
    \end{equation}
    In this equation, $T(t,\lambda)$ represents either $T^F$ or $T^A$ at time $t$ and tempo $\lambda$, while $\rho$ serves as a reference tempo (e.g., 60\,BPM). The variable $s$ indexes the position within the octave, and the sum runs over the equivalence class $[s\cdot\rho]$.
    
    Finally, we extract our 64-dimensional feature set from these representations using the following metrics\footnote{The last two metrics explicitly use non-cyclic peaks to ensure groove-level characteristics are preserved without octave masking.}:
    \begin{itemize}
    	\item \textbf{Statistical Summaries:} Extracted from the top tempo bins of the cyclic tempograms, including their BPM, mean magnitude, and temporal variability.
    	\item \textbf{Temporal Energy Ratio:} Computed from the non-cyclic raw peaks to distinguish half-time from full-time groove density.
    	\item \textbf{Fourier Peak Concentration Index:} Computed from the raw peaks to separate rhythmically straight patterns from highly syncopated ones. 
    \end{itemize}

	To isolate EDM's inherent acoustic organization from potential feature engineering artifacts, we additionally extract embeddings from two state-of-the-art pre-trained models:

	\textbf{MERT-95M} (768 dimensions): a music-specific transformer trained on 160M music clips using masked language modeling~\cite{li2024mert}. MERT learns hierarchical music representations without genre labels.

	\textbf{CLAP-tiny} (512 dimensions): a contrastive audio-language model trained on 630K audio-text pairs~\cite{wu2023clap}. CLAP's multimodal training provides a domain-agnostic perspective.

	These embeddings serve as independent validation: if diverse feature representations converge to similar natural cluster counts despite different architectures, training objectives, and dimensionalities, this provides strong evidence that our discovered taxonomy reflects EDM's inherent acoustic organization rather than methodological choices.

	\subsection{Feature Engineering and Selection}\label{ssec:selection}

	We apply multi-scale preprocessing to enrich the feature space: polynomial expansion (degree-2 interaction terms between features), log-transformation of skewed distributions, and cross-domain products (e.g., spectral $\times$ rhythmic). To stabilize variance and address the severe skewness of many acoustic features, we apply a Yeo--Johnson Power Transformation~\cite{yeo2000new} to approximate Gaussian distributions, optimizing the space for variance-based clustering algorithms like K-Means.

	The expanded feature space (395 features) contains substantial redundancy. We first remove highly correlated features (Pearson $|r| > 0.95$), reducing the set to 183. To evaluate the taxonomy under best-case conditions, we rank the remaining features using a supervised ensemble score ($S_{\mathrm{ens}}$) that averages six label-dependent metrics (including ANOVA F-statistic, Mutual Information, and Random Forest importance) to maximize commercial genre discrimination. This intentionally gives the commercial taxonomy its best chance of being recoverable---if clusters still fail to match genre labels under these favorable conditions, the mismatch cannot be attributed to poor feature choice. However, to prevent this supervised \emph{ranking} from dictating the unsupervised \emph{structure}, the final feature budget $n$ is determined via internal clustering quality. Sweeping budgets $n \in [15, 50]$, we select the top $n=20$ features because this subset maximizes the global silhouette score (Section~\ref{ssec:natural}). This ensures that while feature ranking leverages label information, the structural decisions (feature budget and cluster count) remain label-independent.

	\section{Methods}\label{sec:method}
	\subsection{Clustering Methods}
	We employ two clustering paradigms: divisive hierarchical clustering provides a top-down partitioning approach, while K-means serves as a highly robust, centroid-based optimization.

	\textbf{Hierarchical Divisive Clustering.} We implement a divisive hierarchical clustering algorithm that iteratively splits clusters based on a heterogeneity score:
	\begin{equation}
		H(C) = \sigma^2(C) \cdot (1 + \text{sil}(C)) \cdot \log(|C| + 1)
	\end{equation}
	where $\sigma^2(C)$ represents cluster variance and $\text{sil}(C)$ is the mean silhouette coefficient obtained by tentatively splitting $C$ into two sub-clusters via K-means.

	\textbf{K-means Clustering.} We use K-means++ initialization~\cite{arthur2007kmeans} with 50 restarts to ensure convergence to a stable solution. The algorithm minimizes within-cluster sum of squares. For natural cluster discovery, we test $k \in [15, 30]$ and select optimal $k$ via equal-weight consensus of four validity criteria: elbow method on inertia, silhouette coefficient~\cite{rousseeuw1987}, Calinski-Harabasz index~\cite{calinski1974dendrite}, and Davies-Bouldin index~\cite{davies1979}. Each criterion votes for the $k$ at its elbow point; ties favor the more parsimonious solution.

	\subsection{Evaluation Metrics}\label{ssec:metrics}

	We evaluate clustering through three metric categories: external metrics measure agreement with commercial labels, internal metrics assess acoustic coherence, and distribution metrics prevent trivial solutions.

	\textbf{External Metrics} (requiring ground-truth labels): Normalized Mutual Information (NMI) measures information shared between predicted and true labels; Adjusted Rand Index (ARI) quantifies pairwise agreement corrected for chance; Purity assigns each cluster to its majority class and computes overall accuracy; Mutual Information (MI) Score captures statistical dependence between clusterings.

	\textbf{Internal Metrics} (based on structure): Silhouette coefficient~\cite{rousseeuw1987} measures how well samples fit within their clusters; Davies-Bouldin index~\cite{davies1979} evaluates the ratio of within-cluster to between-cluster distances (lower is better); Cophenetic Correlation Coefficient~\cite{sokal1962} assesses structure preservation through bootstrap stability analysis.

	\textbf{Distribution Metrics}: Cluster Balance (entropy) and Normalized Cluster Balance measure uniformity of cluster sizes, important to avoid trivial solutions.

	\section{Results}\label{sec:results}

	\subsection{Acoustic Mismatch at the Commercial Scale}\label{ssec:constrained}

    	\begin{table}[t]
		\centering
		\caption{Clustering Performance Metrics ($k$ = 35)}
		\label{tab:k35}
		\resizebox{\columnwidth}{!}{%
			\begin{tabular}{l|cc|cc}
				\toprule
				& \multicolumn{2}{c|}{\textbf{Extracted Features}} & \multicolumn{2}{c}{\textbf{*Pre-trained Embeddings}} \\
				\textbf{Metric} & \textsc{K-means} & \textsc{Divisive} & \textsc{MERT-95M} & \textsc{CLAP-tiny} \\
				\midrule
				\multicolumn{5}{l}{\textit{Internal}} \\
				Silhouette$\uparrow$ & \textbf{0.0956} & \underline{0.0666} & 0.0491 & 0.0496 \\
				Davies-Bouldin$\downarrow$ & \textbf{2.0307} & \underline{2.3475} & 2.7440 & 2.7764 \\
				Cophenetic$\uparrow$ & \textbf{0.5840} & \textbf{0.5840} & \underline{0.3657} & 0.2682 \\
				\midrule
				\multicolumn{5}{l}{\textit{External}} \\
				ARI$\uparrow$ & \textbf{0.1487} & 0.0890 & 0.0982 & \underline{0.1174} \\
				NMI$\uparrow$ & \textbf{0.3887} & \underline{0.3206} & 0.2949 & 0.3197 \\
				MI Score$\uparrow$ & \textbf{1.3619} & 1.0912 & 1.0182 & \underline{1.1267} \\
				Purity$\uparrow$ & \textbf{0.3323} & 0.2543 & 0.2177 & \underline{0.2632} \\
				\midrule
				\multicolumn{5}{l}{\textit{Distribution}} \\
				Balance$\uparrow$ & \underline{3.4512} & 3.2521 & 3.3499 & \textbf{3.4940} \\
				Norm.\ Bal.$\uparrow$ & \underline{0.9707} & 0.9147 & 0.9422 & \textbf{0.9827} \\
				\bottomrule
				\multicolumn{5}{l}{\scriptsize $\uparrow$Higher is better, $\downarrow$Lower is better, \textbf{bold} = best, \underline{underline} = 2nd} \\
				\multicolumn{5}{l}{\scriptsize *All pre-trained embeddings use K-means clustering} \\
			\end{tabular}%
		}
	\end{table}
    
	Table~\ref{tab:k35} compares clustering performance at $k$=35---matching the number of commercial genres---across our extracted features and two pre-trained embedding baselines. No method produces meaningful alignment with commercial labels, with even the best NMI and ARI falling well below accepted thresholds for cluster-label correspondence (NMI$>$0.6, ARI$>$0.3). This uniformly poor agreement is itself a central finding---it indicates a systematic mismatch between acoustic reality and commercial taxonomy, independent of feature representation or clustering algorithm.

	Within this landscape, our handcrafted features outperform pre-trained embeddings on internal quality, achieving roughly double the silhouette scores and $25\%$ lower Davies-Bouldin indices than both MERT and CLAP. This gap suggests that EDM genre structure is better captured by domain-specific temporal and production descriptors than by general-purpose audio representations. Among our methods, K-means produces higher-quality clusters than Divisive across all metrics---consistent with the silhouette-optimized selection and Yeo--Johnson normalization, which create a feature space geometrically suited to centroid-based partitioning.

	Notably, CLAP achieves marginally better cluster balance than our features, likely encoding label-like semantic structure through its audio-language training that distributes tracks more evenly---but less precisely---across clusters. This trade-off reinforces our thesis: maximizing acoustic separability does not recover the commercial taxonomy, because many genre boundaries are drawn along non-acoustic lines.

	\begin{figure}[!t]
		\centering
		\includegraphics[width=\columnwidth]{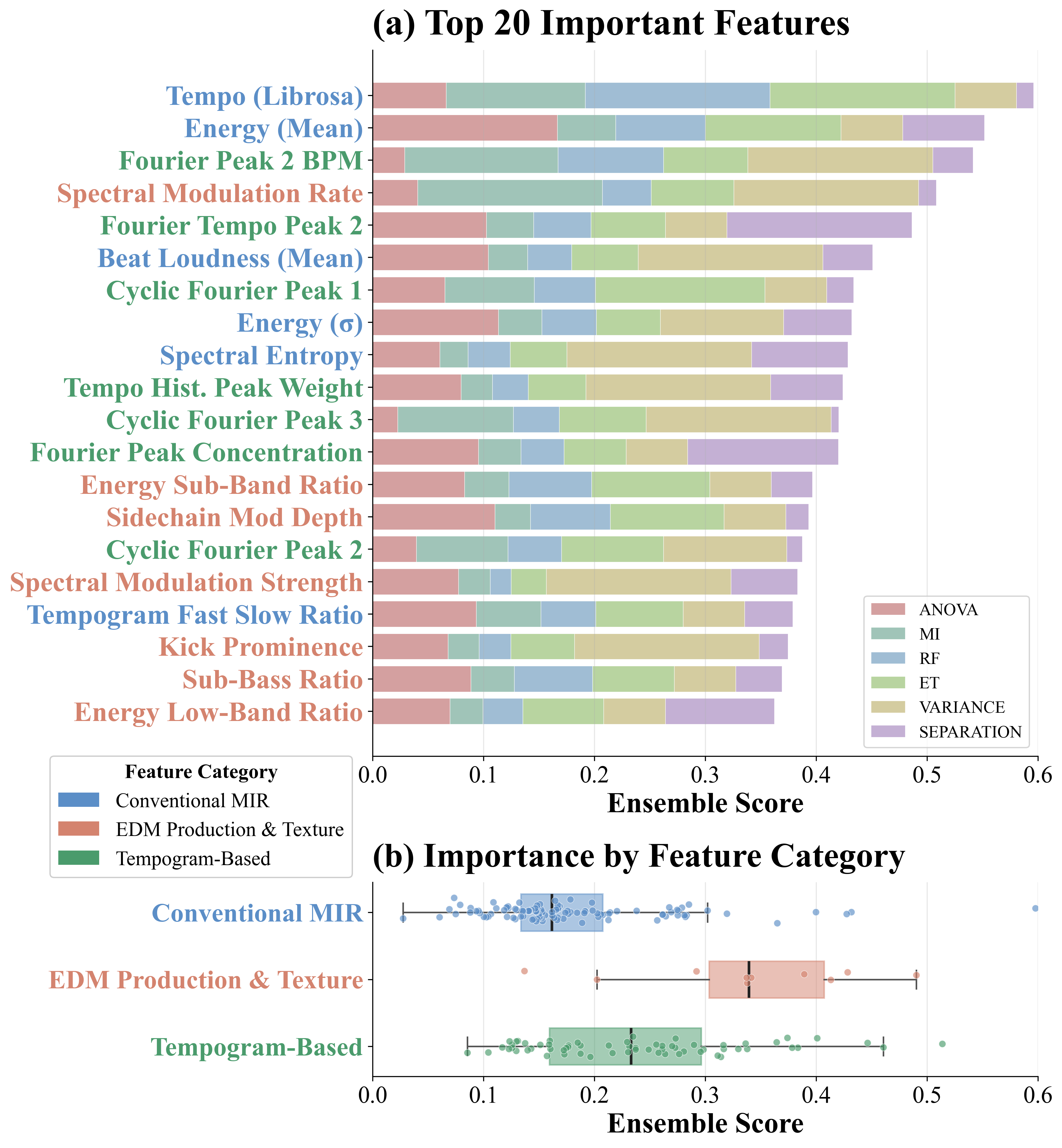}
		\caption{Feature importance analysis. (a) Top 20 features ranked by supervised ensemble score. (b) Importance distributions grouped by feature category: EDM-specific (Production \& Texture, Tempogram-based) versus conventional MIR descriptors.}
		\label{fig:features}
	\end{figure}
    
	To understand which acoustic dimensions drive commercial genre distinctions, we evaluate the discriminative power of our feature space. Tempo and tempogram descriptors rank highly alongside energy and spectral modulation features (Figure~\ref{fig:features}), and EDM-specific categories (Production \& Texture, Tempogram-based) show elevated median importance relative to conventional MIR timbral descriptors. This suggests that genre identity in EDM draws on a broad mix of temporal, energetic, and timbral cues, with production aesthetics and rhythmic structure contributing alongside static spectral properties.

	\subsection{Natural Clustering Structure}\label{ssec:natural}

	Since K-means yields the highest-quality clusters at $k$=35 and the consensus criteria for optimal-$k$ selection require sweeping over $k$ values---a procedure native to partitional methods---we adopt K-means for natural-$k$ discovery. When removing the $k$=35 constraint, three of four validity criteria (silhouette, inertia elbow, Calinski-Harabasz; see the supplementary material\footnote{See Appendix~B in the supplementary material.} for detailed curves) identify optimal cluster counts between 17 and 20, with only Davies-Bouldin selecting a higher value ($k$=29). When sweeping the feature budget across $n \in [15, 50]$, the global structural maximum (peak internal silhouette score) aligns with this range at $n=20$ features. This structural robustness supports our core thesis: even within a feature space optimized to maximally discriminate the commercial labels, unsupervised geometric optimization consistently rejects the 35-genre dimensionality in favor of 17--20 coherent groups---roughly 40--50\% fewer than the prescribed taxonomy.

	We emphasize that this finding concerns \emph{acoustic} overspecification and does not imply that the commercial taxonomy is inherently flawed. Genre labels serve cultural, historical, and commercial functions that extend well beyond acoustic separability---they encode community identity, regional scene affiliations, DJ set-programming conventions, and listener expectations shaped by years of cultural practice~\cite{brackett2016}. Two genres may be acoustically near-identical yet culturally distinct, and such distinctions are legitimate from a music-industry perspective---Appendix~D of the supplementary material\footnote{See Appendix~D in the supplementary material, available at \url{https://github.com/alunxu/edm-overspec}.} illustrates several such ``acoustically convergent'' pairs, whose radar profiles nearly overlap despite carrying different commercial labels. Our claim is narrower: when evaluated on the basis of audio features, 40--50\% of the prescribed 35 categories lack sufficient acoustic differentiation to form distinct clusters, suggesting that the current taxonomy subdivides what are, acoustically, coherent styles.

	\subsubsection{Acoustic Topography}

    \begin{figure*}[t]
		\centering
		\includegraphics[width=\textwidth]{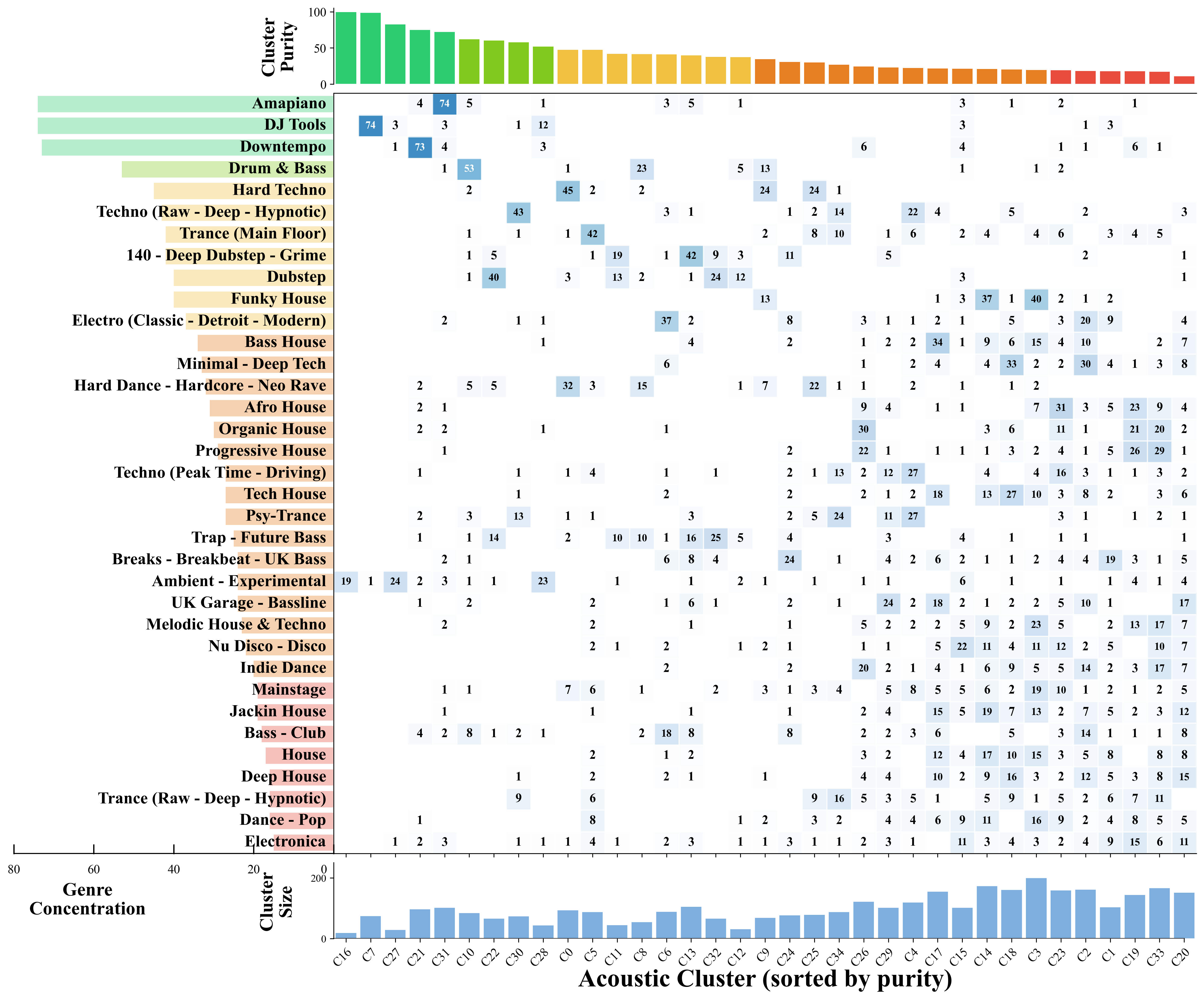}
		\caption{Confusion matrix mapping 35 commercial genre labels to 35 discovered acoustic clusters. The matrix is augmented by three marginals: \textit{Cluster Purity} (top) measures the acoustic exclusivity of a cluster; \textit{Genre Concentration} (left) measures how acoustically monolithic a commercial genre is; and \textit{Cluster Size} (bottom) indicates the total volume of tracks in that acoustic space. The clear inverse relationship between cluster size and purity highlights how massive, generic acoustic templates absorb multiple overlapping commercial labels, while smaller, pure clusters represent distinct sonic niches.}
		\label{fig:confusion}
	\end{figure*}

	Figure~\ref{fig:confusion} visualizes the mapping between commercial labels and acoustic clusters, augmented with three structural marginals: Cluster Purity, Cluster Size, and Genre Concentration. Analyzing these marginals reveals the complex topography of EDM, which transitions from distinct, isolated acoustic niches to generic stylistic melting pots.

	At one extreme, clusters with both high purity and high genre concentration---such as those dominated by Amapiano, DJ Tools, Downtempo, and Drum \& Bass---indicate cases where the commercial taxonomy closely mirrors acoustic reality. These styles are acoustically monolithic: producers adhere to a narrow set of defining sonic rules, and the resulting clusters form isolated ``crystallized'' islands in the feature space that resist hybridization.

	Conversely, the largest clusters (bottom margin) consistently exhibit the lowest purity (top margin). These massive acoustic hubs absorb tracks from heavily overlapping commercial categories, acting as ``liquid zones.'' The inverse relationship between cluster size and purity implies a shared mainstream acoustic template: the taxonomy assigns multiple labels to what is acoustically a shared sonic space, even though the tracks are indistinguishable at a structural level.

	A third pattern emerges among genres with low concentration (left margin), whose tracks are dispersed across numerous distinct acoustic clusters. Broad or rapidly evolving labels often fail to localize into a single acoustic signature; the commercial label in these cases functions as a cultural umbrella for diverse production styles that our clustering separates. The interplay of these three marginals demonstrates that overspecification is not uniform: it is concentrated in the mainstream center, while the stylistic periphery remains acoustically crisp.

	\subsubsection{Genre Purity and Acoustic Extremity} \label{ssec:profile}

	\begin{figure}[t]
		\centering
		\includegraphics[width=0.95\columnwidth]{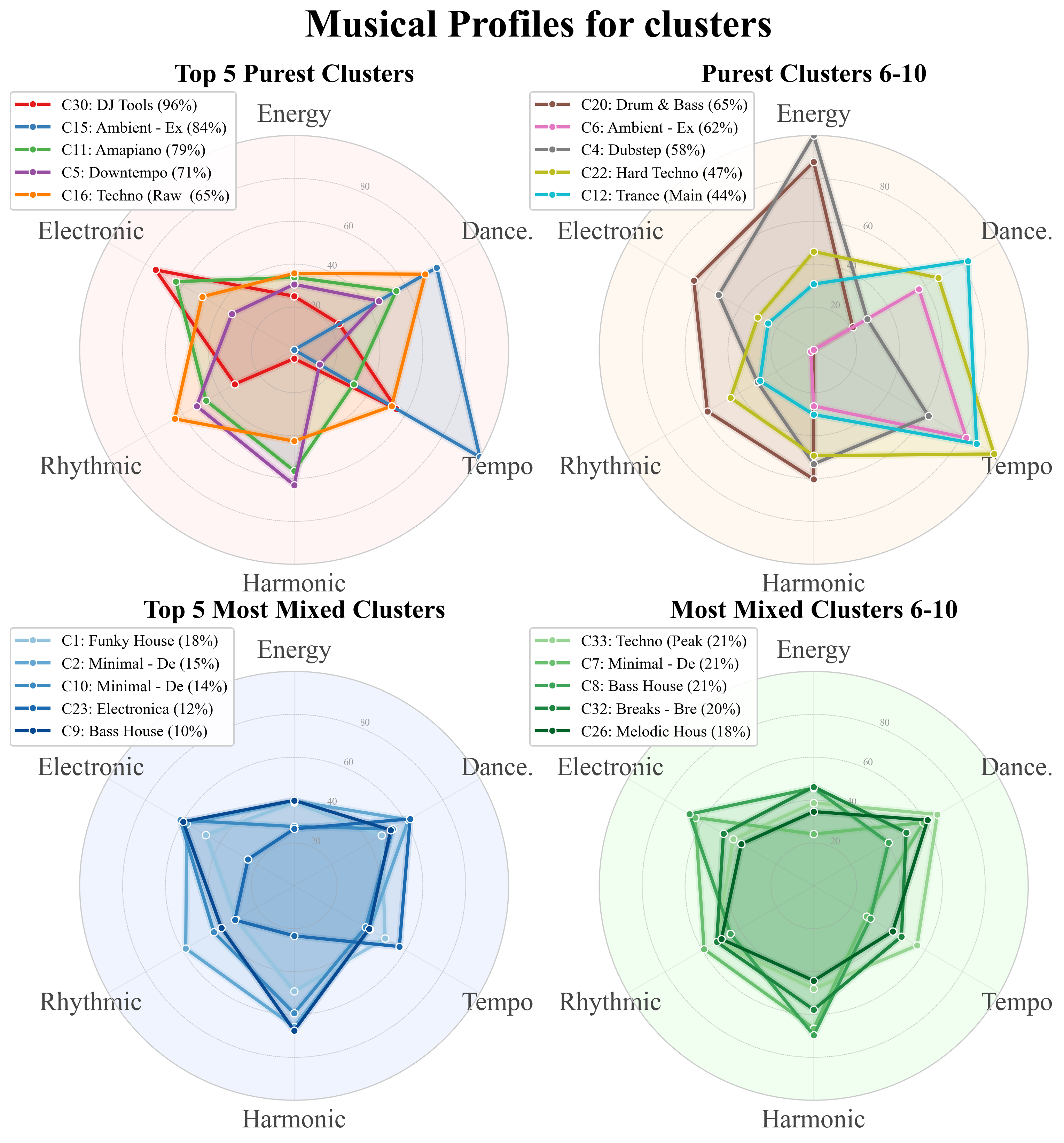}
		\caption{Music profiles of selected clusters, comparing the five purest (top) against the five most hybridized (bottom). Each axis summarizes one of six acoustic dimensions---Energy, Danceability, Tempo, Harmonic complexity, Rhythmic density, and Electronic texture---by z-score normalizing all constituent features within a dimension, averaging them per track, and rescaling cluster means to 0--100 via $p_5$/$p_{95}$ interpolation. Pure clusters exhibit extreme, peaked profiles; mixed clusters converge toward moderate, overlapping shapes.}
		\label{fig:profiles}
	\end{figure}

	To summarize each cluster's acoustic signature, we assign every extracted feature to one of six interpretable dimensions---Energy, Danceability, Tempo, Harmonic complexity, Rhythmic density, and Electronic texture---via an explicit, manually curated mapping that covers all features in the dataset (Section~\ref{ssec:features}), including the EDM-specific production descriptors. Within each dimension, the constituent features span different physical units and scales (e.g., decibels vs.\ spectral ratios); we therefore z-score normalize each feature independently across the full corpus so that no single feature dominates the aggregate. The normalized values are then averaged across all features in that dimension to yield a single per-track score. Finally, to map cluster-level means onto an interpretable 0--100 scale while remaining robust to outliers, we linearly interpolate between the 5th and 95th percentiles (\mbox{$p_5$/$p_{95}$}) of the corpus-wide per-dimension distribution and clip to $[0,100]$.

	Figure~\ref{fig:profiles} reveals that genre purity correlates with acoustic extremity. The clusters with highest purity occupy feature-space vertices: DJ Tools---a utility category of loops, samples, and mixing elements---maximizes Electronic texture while scoring low on Energy, Danceability, and Rhythmic dimensions, and its tracks span multiple clusters because the category encompasses diverse production styles. Similarly, Ambient-dominated clusters push Harmonic features to extremes while compressing Rhythmic and Energy dimensions. We additionally note that the ``Purest Clusters 6--10'' panel identifies Hard Techno (47\%) as relatively pure. While acoustically defensible here, this conflicts with the recent expansion of ``Hard Techno'' on Beatport, where the category has become heterogeneous (incorporating elements from industrial, acid, and trance). The 47\% purity might be inflated by the March 2025 snapshot capturing a more codified moment, but it is worth noting that this genre's boundaries are in rapid flux. These poles---whether functional (DJ Tools), artistic (Ambient), or temporally codified (Hard Techno)---demonstrate how specialized production constraints create acoustic signatures that resist genre hybridization.

	Conversely, the most mixed clusters exhibit moderate, overlapping profiles rather than converging to a single ``neutral'' center. Notably, certain genres recur across multiple mixed clusters at different purity levels---Bass House and Minimal--Deep Tech each appear in several low-purity clusters, indicating that these labels fragment across acoustically distinct regions of the feature space. Each hybrid style draws on parent traditions differently---bass-heavy clusters extend along Rhythmic and Electronic axes, while groove-oriented clusters balance Danceability and Energy---suggesting that genre blending follows multiple distinct pathways through the feature space.

	\subsubsection{Implications for Genre Organization}

	The pattern observed in Figure~\ref{fig:profiles} suggests EDM operates through two distinct modes: ``crystallized'' genres at acoustic extremes maintaining distinct identities through specialized production choices, and ``liquid'' zones in the center where hybridized production styles dynamically overlap. The improvement in clustering quality when allowing natural groupings ($k$=17--20 vs $k$=35) combined with these purity patterns indicates that many commercial genre distinctions may subdivide what are acoustically coherent styles, while genuine acoustic boundaries occur at the extremes where production techniques diverge most dramatically---a core-periphery structure that any genre taxonomy or retrieval system should account for.

	\begin{figure}[!t]
		\centering
		\includegraphics[width=0.45\textwidth]{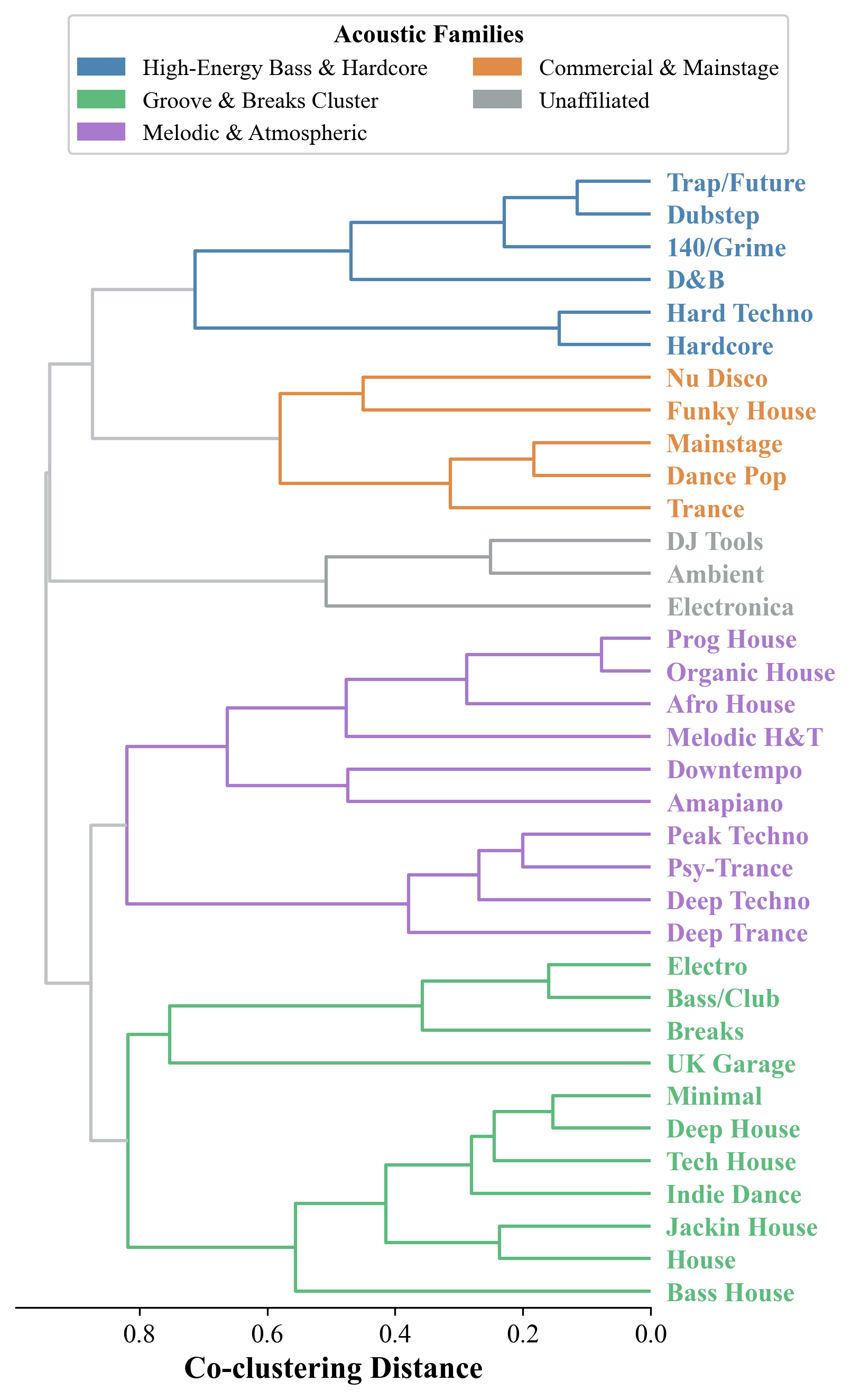}
		\caption{Hierarchical genre lineage derived from co-clustering affinity ($1-\hat{a}_{ij}$, where $\hat{a}_{ij}$ is the normalized co-clustering frequency across 50 K-means restarts). Four acoustic families and three isolated outliers emerge.}
		\label{fig:dendrograms}
	\end{figure}
    
	\subsection{Hierarchical Genre Structure}\label{ssec:hierarchy}

	To reveal hierarchical structure, we compute a \emph{co-clustering affinity} between every genre pair: across 50 independent K-means runs ($k$=35), we record for each genre pair $(i,j)$ the fraction of their tracks that co-occur in the same cluster, yielding a symmetric affinity $\hat{a}_{ij} \in [0,1]$. Converting affinity to distance ($1-\hat{a}_{ij}$) and applying average linkage yields the dendrogram in Figure~\ref{fig:dendrograms}. The resulting tree reveals that EDM's acoustic space is organized not as a set of discrete genre families but as a gradient between two poles, separated by a broad middle zone where most commercial labels overlap.

	At one pole, Drum \& Bass, Dubstep, 140/Grime, Hard Techno, and Hardcore merge internally before joining any other branch, forming the most cohesive family in the tree. This cohesion is consistent with the acoustic dimension of Reynolds' ``hardcore continuum'' thesis~\cite{reynolds2013energy}: shared production techniques---chopped breaks, sub-bass emphasis, and high-energy distortion---create a signature that resists mainstream hybridization. Notably, Hard Techno clusters here despite its four-on-the-floor kick pattern, revealing that tempo and spectral intensity dominate rhythmic patterns in determining acoustic affinity at the extremes: its 145+ BPM tempos and industrial distortion place it closer to Hardcore than to the mid-tempo techno variants with which it shares cultural lineage. Similarly, Trap decouples from full-time Drum \& Bass and clusters with half-time Dubstep, validating the temporal energy ratio descriptor.

	At the opposite extreme, Ambient, DJ Tools, and Electronica merge last, occupying isolated positions defined by specialized production constraints that place them at feature-space vertices far from the mainstream.

	Between these poles, the majority of commercial labels interleave into three families within a densely connected middle zone: a large \emph{Groove \& Breaks} cluster (the House family, Electro, Breaks, UK Garage); a \emph{Melodic \& Atmospheric} gradient (Peak Techno, Psy-Trance, Afro House, Organic House, Downtempo); and a \emph{Commercial \& Mainstage} cluster (Dance Pop, Mainstage, Nu Disco, Funky House, Trance). Two placements merit comment: Nu Disco and Funky House join the Commercial branch rather than the Groove cluster, suggesting that shared vocal hooks and major-key tonality outweigh groove-level similarity; and main-floor Trance's proximity to Mainstage likely reflects convergent production aesthetics (euphoric builds, layered pads, anthemic structure). Within this gradient, the rhythm-aware descriptors separate syncopated from straight-kick genres---Peak Techno groups with Psy-Trance in the Melodic branch, while Breaks and UK Garage map to the Groove cluster---yet labels that fragment this middle zone (Tech House vs.\ Minimal, Progressive House vs.\ Melodic House \& Techno) still collapse structurally, suggesting they subdivide overlapping styles for cultural rather than acoustic reasons.

	\section{Conclusion}\label{sec:conclusion}
	This study examined the acoustic basis of EDM's commercial taxonomy by clustering 3,500 Beatport tracks across 35 genres. The results indicate acoustic overspecification: natural clustering consistently identifies 17--20 distinct groups---roughly 40--50\% fewer than prescribed. This reduction remains robust even when features are selected to maximize genre-class discrimination, and convergent evidence from pre-trained embeddings (MERT, CLAP) corroborates the structural collapse. Genre categories emerge from artist and community practice before being codified into fixed taxonomies; our results suggest this codification preserves more distinctions than the acoustic signal warrants---raising the hypothesis, to be tested in other genres and platforms, that genre labels tend toward overspecification when cultural and acoustic criteria are conflated.

	Two limitations qualify these findings. First, our analysis is purely acoustic and does not capture the cultural, historical, or scene-level factors that legitimately motivate genre distinctions. Second, many EDM genres differ through subtle timbral cues---drum-machine palettes, synthesis techniques, groove micro-timing---that may not be fully resolved by our feature set; some genre merging may therefore reflect insufficient feature granularity rather than genuine acoustic equivalence. Future work should investigate whether fine-grained timbral descriptors and cultural metadata can recover distinctions that our current features miss.

	\begin{acknowledgments}
		The authors would like to thank Jingbang Liu for developing the web scraping infrastructure used for systematic data collection from Beatport, and Haotian Fang for productive discussions regarding feature selection. Their contributions were instrumental to this research.
	\end{acknowledgments}

	\bibliography{smc2026bib}

\end{document}